\documentclass[a4paper,12pt]{article}
\usepackage{amsmath}
\usepackage{amssymb}
\usepackage{latexsym}
\usepackage{dsfont}
\newcommand{\be}{\begin{eqnarray}}
\newcommand{\ee}{\end{eqnarray}}

\newcommand{\ph}{\phantom}
\newcommand{\bs}{\boldsymbol}
\newcommand{\bsub}{\begin{subequations}}
\newcommand{\esub}{\end{subequations}}
\newcommand{\sech}{\textrm{sech}}
\newcommand{\arccosh}{\textrm{arccosh}}

\begin{document}
\title{\large\textbf{The General Solution of Bianchi Type III Vacuum Cosmology}}
\author{\textbf{T. Christodoulakis}\thanks{tchris@phys.uoa.gr} ~\textbf{and}~\textbf{Petros A. Terzis}\thanks{pterzis@phys.uoa.gr}\\ University of Athens, Physics Department\\
Nuclear \& Particle Physics Section\\
Panepistimioupolis, Ilisia GR 157--71, Athens, Hellas}
\date{}
\maketitle
\begin{center}
\textit{}
\end{center}
\vspace{0.5cm} \numberwithin{equation}{section}
\begin{abstract}
The second order Ordinary Differential Equation which describes the
unknown part of the solution space of some vacuum Bianchi
Cosmologies is completely integrated for Type III, thus obtaining
the general solution to Einstein's Field Equations  for this case,
with the aid of the sixth Painlev\'{e} transcendent $P_{VI}$.  For
particular representations of $P_{VI}$ we obtain the known
Kinnersley two-parameter space-time and a solution of Euclidean
signature. The imposition of the space-time generalization of a
"hidden" symmetry of the generic Type III spatial slice, enables us
to retrieve the two-parameter subfamily without considering the
Painlev\'{e} transcendent.
\end{abstract}
\newpage
\section{Introduction}
In a recent work of ours \cite{ChrTer}, the theory of symmetries of
systems of coupled, ordinary differential equations (ODE's) has been
used to develop a concise algorithm for cartographing the space of
solutions to vacuum Bianchi Einstein's Field Equations (EFE). The
symmetries used were the well known automorphisms of the Lie algebra
for the corresponding isometry group of each Bianchi Type, as well
as the scaling and the time reparameterization symmetry. Application
of the method to Type III resulted in: a) the recovery of most known
solutions without prior assumption of any extra symmetry, b) the
enclosure of the entire unknown part of the solution space into a
single, second order ODE in terms of one dependent variable and c) a
partial solution to this ODE. It is worth-mentioning the fact that
the solution space were thus seen to be naturally partitioned into
three distinct, disconnected pieces: one consisting of the known
Siklos (pp-wave) solution \cite{Siklos}, another occupied by the
Type III member of the known Ellis-MacCallum family
\cite{MacBook},\cite{Wainr} and the third described by the
aforementioned ODE. Lastly, preliminary results reported have shown
that the unknown part of the solution space for other Bianchi Types
is described by a strikingly similar ODE, pointing to a natural
operational unification, at least as far as the problem of solving
the cosmological EFE's is concerned. In section 2 of this work we
present the general solution to the ODE in question for  the case of
Type III, that is the general Type III vacuum Geometry. The metric
components are implicitly given in terms of the sixth Painlev\'{e}
transcendent. To overcome this practical difficulty we use the
existing knowledge on particular, truly closed form (i.e. in terms
of elementary functions of time) representations of the $P_{VI}$
transcendent; work on finding such solutions can be found in the
contributions to a recent meeting at the Newton Institute
\cite{Newton}, and also in \cite{Okamoto}, \cite{KirTan}. For other
cases of Painlev\'{e} solutions in relativity see, e.g.
\cite{Wills}, \cite{Calvert}, \cite{Manoj} and (13.60), (13.70) in
\cite{MacBook}. For a recent account on how Painlev\'{e}
transcendents occur in dimensional reductions of integrable systems
see \cite{Mason}. This investigation results in the recovery of the
known Kinnersley \cite{Kinner} two parameter family of metrics,
admitting a $G_{4}$ multiply transitive isometry group and a
bi-parametric solution of Euclidean signature, admitting only the
initial $G_{3}$ isometry. In section 3, the existence of an extra
inherent symmetry of the general Type III 3-geometry is exploited in
order to arrive at the aforementioned Kinnersley solution by a prior
assumption of extra symmetry. Finally, some concluding remarks are
included in section 4, while some mathematical aspects of the
derivation in section 3 are described in the Appendix.

\section{The General Solution}

As it is well known, for spatially homogeneous space-times with a
simply transitive action of the corresponding isometry group
\cite{MacBook}, \cite{EM}, the line element assumes the form
\begin{equation}\label{line element}
ds^2=\left(N^\alpha N_\alpha-N^2\right)dt^2+2\,N_\alpha\,
\sigma^\alpha_{\ph{a}i}\,dx^i\,dt+
\gamma_{\alpha\beta}\,\sigma^\alpha_{\ph{a}i}\,\sigma^\beta_{\ph{a}j}\,dx^i\,dx^j
\end{equation}
where the 1-forms $\sigma^{\alpha}_{\ph{a}i}$, are defined from:
\begin{equation}\label{ορισμός σ}
d\sigma^{\alpha}=C^\alpha_{\ph{a}\beta\gamma}\,\sigma^\beta\wedge\sigma^\gamma
\Leftrightarrow \sigma^{\alpha}_{\ph{a}i,\,j} -
\sigma^\alpha_{\ph{a}j,\,i}=2\,C^\alpha_{\ph{a}\beta\gamma}\,
\sigma^{\gamma}_{\ph{a}i} \sigma^\beta_{\ph{a}j}.
\end{equation}

( small Latin letters denote world space indices while small Greek
letters count the different basis one-forms; both type of indices
range over the values 1,2,3 )

 For Bianhi Type III Cosmology the structures constants are
\cite{Ryan}
\begin{equation}\label{σταθερές δομής}
\begin{array}{ll}
C^1_{\ph{1}13}=-C^1_{\ph{1}31}=1\\
C^\alpha_{\ph{a}\beta\gamma}=0 & for\, all\, other\, values\, of\,
\alpha \beta \gamma
\end{array}
\end{equation}
Using these values in the defining relation (\ref{ορισμός σ}) of
the 1-forms $\sigma^{\alpha}_{\ph{a}i}$ we obtain
\begin{equation}
\sigma^\alpha_{\ph{a}i}=\left(\begin{matrix} 0 & e^{-x} & 0 \cr 0
& 0 & 1 \cr \frac{1}{2} & 0 & 0
\end{matrix}
\right)
\end{equation}
The corresponding vector fields $\xi^i_{\ph{a}\alpha}$ (satisfying
$\left[\xi_\alpha,\xi_\beta\right]=C^\gamma_{\ph{a}\alpha\beta}\,
\xi_\gamma$) with respect to which the Lie Derivative of the above
1-forms is zero are: \be\label{killing}
\begin{array}{lll}
\xi_1=\partial_y & \xi_2=\partial_z & \xi_3=\partial_x+y\partial_y
\end{array}
\ee In the recent work of ours \cite{ChrTer}, we retrieved  most
known solutions of Bianchi Type III and we showed that the unknown
part of the solution space of this Cosmology is described, without
loss of generality, by a line element of the form:

\be\label{lin el Na=0} ds^2=-N^2\,d\rho^2+
\gamma_{\alpha\beta}\,\sigma^\alpha_{\ph{a}i}\,\sigma^\beta_{\ph{a}j}\,dx^i\,dx^j
\ee where the scale factor matrix $\gamma_{\alpha\beta}(\rho)$ and
the lapse function $N(\rho)$ are given by the equations:
\be\label{gamma u}
(N)^2=\frac{{u'}^2-1}{8\,(3\,\rho-5)\,(\rho^2-u^2-1)}\,e^{u_1} \quad
\text{and} \quad \gamma_{\alpha\beta}=\left(
\begin{matrix}
e^{{u_1} + 2\,{u_4}} & e^{{u_1} + {u_2} + {u_4}} & 0 \cr e^{{u_1}
+ {u_2} + {u_4}} & {\frac{3\,\rho-3 }{ 3\,\rho -5}}\,e^{{u_1} +
2\,{u_2}} & 0 \cr 0 & 0 & e^{{u_1}}
\end{matrix}
\right) \ee

The functions $u_1,u_2,u_4$ satisfy \be u_1'& = &\frac{ -3\,u +
\left( 3\,\rho-1  \right) \,u' }{2\,\left( {u'}^2-1 \right) }\,
      u''\\
      u_2' & = & \frac{ \left( 3\,\rho-1  \right) \,
      \left( -1 + {u'}^2 + {\left(  3\,\rho-5  \right) }^2\,u\,u''
      -
        {\left( 3\,\rho-5  \right) }^2\,\left(  \rho-1  \right) \,u'\,u''
        \right)   }{4\,{\left( 3\,\rho-5  \right) }^2\,\left(  \rho-1  \right) \,
    \left(  {u'}^2-1 \right) }\\
    u_4' & = & \frac{3\,\rho-5}{3\,\rho-1}\,u_2' \ee
and the function $u(\rho)$ obeys a second order differential
equation, of the form:

\be\label{final u III} \ddot{u}^2=\frac{(-1+\dot{u}^2)^2}
{(\kappa+\lambda\,\rho)\,(\rho^2-u^2-1)} & \kappa=-10,\,\lambda=6
\ee

In order to solve (\ref{final u III}), for arbitrary constants
$(\kappa,\,\lambda)$, we apply the contact transformation:
\be\label{contact}
\begin{split}
u(\rho)& =
-\frac{8}{\lambda}\,y(x)+\frac{4\,(2x-1)}{\lambda}\,y'(x) &\rho & =  -\frac{\kappa}{\lambda}+\frac{4}{\lambda}\,y'(x)\\
u'(\rho)& =  2\,x-1 & u''(\rho)& = \frac{\lambda}{2\,y''(x)}
\end{split}\ee
which reduces it to \be\label{y equation} x^2\,(x-1)^2\,{y''}^2=
-4y'\,(x\,y'-y)^2+4\,{y'}^2\,(x\,y'-y)-\frac{\kappa}{2}\,{y'}^2+
\frac{\kappa^2-\lambda^2}{16}\,y'
 \ee

This equation is a special form of the equation SD-Ia appearing in
\cite{Cosgrove}, a work in which a classification of second order,
second degree ordinary differential equations has been performed.
The general solution to (\ref{y equation}) is obtained with the help
of the sixth Painlev\'{e} transcendent
$P:=\mathbf{P_{VI}}(\alpha,\beta,\gamma,\delta)$ and reads:
\be\label{solution y} y & = &
\frac{x^2\,(x-1)^2}{4\,P\,(P-1)(P-x)}\,\left(P'-\frac{P\,(P-1)}{x\,(x-1)}\right)^2
\nonumber \\ & & +\frac{1}{8}\,(1\pm
\sqrt{2\,\alpha})^2\,(1-2\,P)-\frac{\beta}{4}\,\left(1-\frac{2\,x}{P}\right)
\nonumber \\ & &
-\frac{\gamma}{4}\,\left(1-\frac{2\,(x-1)}{P-1}\right)+\left(\frac{1}{8}-\frac{\delta}{4}\right)
\, \left(1-\frac{2\,x\,(P-1)}{P-x}\right) \ee
$P:=\mathbf{P_{VI}}(\alpha,\beta,\gamma,\delta)$ is defined by the
ODE:

\be\label{Painleve 6} P'' & = & \frac{1}{2}\left( \frac{1}{-1 + P} +
\frac{1}{P} + \frac{1}{-x + P} \right) \,{P'}^2 -\left( \frac{1}{-1
+ x} + \frac{1}{x} + \frac{1}{-x + P}
\right) \,P' \nonumber \\
& & +\frac{\left( -1 + P \right) \,P\,\left( -x + P \right) }
  {{\left( -1 + x \right) }^2\,x^2} \left(\alpha  + \frac{\left( -1 + x \right) \,\gamma }
  {{\left( -1 + P \right) }^2} +
  \frac{x\,\beta }{{P}^2} + \frac{\left( -1 + x \right) \,x\,\delta }
   {{\left( -x + P \right) }^2}\right) \ee

 where the values of the parameters
$\left(\alpha,\beta,\gamma,\delta\right)$  must satisfy the
following system: \bsub\label{system} \be \alpha-\beta+\gamma-\delta
\pm
\sqrt{2\,\alpha}+1 & =&-\frac{\kappa}{2} \\
\left(\beta+\gamma\right)\,\left(\alpha+\delta \pm \sqrt{2\,\alpha}\right) &=&0 \\
\left(\gamma-\beta\right)\,\left(\alpha-\delta \pm
\sqrt{2\,\alpha}+1\right)+\frac{1}{4}\,\left(\alpha-\beta-\gamma+\delta
\pm \sqrt{2\,\alpha}\right)^2 & = & \frac{\kappa^2-\lambda^2}{16} \\
\frac{1}{4}\,\left(\gamma-\beta\right)\,\left(\alpha+\delta \pm
\sqrt{2\,\alpha}\right)^2+\frac{1}{4}\,\left(\beta+\gamma\right)^2\,\left(\alpha-\delta
\pm \sqrt{2\,\alpha}+1\right) & = & 0 \ee \esub

If we insert in (\ref{system}) the values  $\kappa=-10, \lambda=6$
for Type III, we have twelve solutions to this system. Eight of them
correspond to the "$-\sqrt{2\,\alpha}$" case and the rest four to
the "$+\sqrt{2\,\alpha}$" case.

\textbf{Case I: $\bs{-\sqrt{2\,\alpha}}$} \bsub\label{- sqrt}
\be\label{-a} \alpha = 0,\, \beta =0, \, \gamma = 0, \, \delta = -4
\\ \label{-b}
\alpha = 0, \,  \beta =-2, \, \gamma = 2, \, \delta = 0 \\
\label{-c} \alpha = 2,\, \beta =0, \, \gamma = 0, \, \delta = -4 \\
\label{-d} \alpha = 2,\, \beta =-2, \, \gamma = 2, \, \delta = 0 \\
\label{-e} \alpha = 8,\, \beta =0, \, \gamma = 0, \, \delta = 0 \\
\label{-f} \alpha = \frac{1}{2}, \, \beta =-\frac{9}{2}, \, \gamma =
\frac{1}{2}, \, \delta = \frac{1}{2}\\ \label{-g} \alpha =
\frac{1}{2}, \, \beta =-\frac{1}{2}, \, \gamma = \frac{9}{2}, \,
\delta = \frac{1}{2}\\ \label{-h} \alpha = \frac{9}{2}, \, \beta
=-\frac{1}{2}, \, \gamma = \frac{1}{2}, \, \delta =-
\frac{3}{2}\ee\esub

 \textbf{Case II: $\bs{+\sqrt{2\,\alpha}}$} \bsub\label{+ sqrt} \be
 \label{+a}
\alpha = 0,\, \beta =0, \, \gamma = 0, \, \delta = -4 \\
\label{+b} \alpha = 0, \,  \beta =-2, \, \gamma = 2, \, \delta = 0
\\ \label{+c} \alpha =
2,\, \beta =0, \, \gamma = 0, \, \delta = 0 \\ \label{+d} \alpha =
\frac{1}{2}, \, \beta =-\frac{1}{2}, \, \gamma = \frac{1}{2}, \,
\delta = -\frac{3}{2}\ee\esub

Of course the first two solutions in (\ref{+ sqrt}) are identical to
the first two in (\ref{- sqrt}) and are only written down for the
sake of completeness.

Despite the fact that (\ref{solution y}) describes the general
solution of Bianchi Type III Vacuum Cosmology, this solution does
not come in a particularly manageable form due to the appearance of
the function $\mathbf{P_{VI}}(\alpha,\beta,\gamma,\delta)$. As
mentioned in the introduction, there is a vast body of literature
concerning $\mathbf{P_{VI}}$. The most effective way to find its
closed form solutions, is to apply the following lemma (often called
"linearisability condition") \cite{TamaR}, \cite{Gromak}:

\textbf{Lemma:} Assume that $P(x)$ satisfies the Riccati equation
\be\label{Riccati}
x\,(x-1)\,P'(x)=a\,P(x)^2+(b\,x+c)\,P(x)-(a+b+c)\,x \ee Then $P(x)$
satisfies: \be\label{RicII}
\bs{P_{VI}}\left(\frac{a^2}{2},-\frac{(a+b+c)^2}{2},\frac{(a+c)^2}{2},\frac{1-(1-a-b)^2}{2}\right)
\ee

The proof is elementary: Solving \eqref{Riccati} and its first
derivative for $P'(x), P''(x)$ and substituting in \eqref{Painleve
6} we get an identity for the specific values of the parameters
$(\alpha,\beta,\gamma,\delta)$ satisfying the values of
\eqref{RicII}.

The application of  this lemma proceeds as follows: For every case
of \eqref{- sqrt} and \eqref{+ sqrt}, we find the values of
$(a,b,c)$ (if they exist). Then we solve the corresponding
\eqref{Riccati}. Subsequently, we insert each solution to
\eqref{solution y}, thereby obtaining $y(x)$. Finally inserting this
$y(x)$ into \eqref{contact} we obtain $(u(\rho),\rho)$ in parametric
form in terms of the time variable $x$. The results of this
procedure are:

\subsection*{Case I: $\bs{\alpha = 0,\, \beta =0, \, \gamma = 0, \, \delta = -4}$}
It cannot be linearized.

\subsection*{Case II: $\bs{\alpha = 0, \,  \beta =-2, \, \gamma = 2, \, \delta =
0}$} We get two solutions of \eqref{Painleve 6} \bsub\be
P(x)&=&\frac{x\,\left( 2 + 2\,x\,\log (x) + c\,x \right) }{{\left(
-1 + x \right) }^2} \\ && \nonumber \\ P(x)&=&\frac{-3 + 4\,x -
2\,{\left( -1 + x \right) }^2\,\log (-1 + x) + c\,{\left( -1 + x
\right) }^2}{x^2}\ee\esub but both of them make the B\"{a}cklund
transformation \eqref{contact} degenerate, since they give $\rho
\rightarrow \frac{5}{3}$.
\subsection*{Case III: $\bs{\alpha = 2,\, \beta =0, \, \gamma = 0, \, \delta = -4}$}
We get one solution \be P(x)=\frac{{\left( -1 + x \right) }^2}{1 -
2\,x + \,c\,x^2 } \ee which makes \eqref{contact} \bsub\label{u 2.3}
\be {{u(\rho )}=
    {\frac{4\,\left( -1 - 2\,\left( -1 + c \right) \,x + 6\,c\,x^2 - 4\,c\,x^3 +
          c\,\left( 2 + c \right) \,x^4 \right) }{3\,{\left( -1 + 2\,x + c\,x^2 \right)
          }^2}}} \\
  {{\rho }=
    {\frac{5 + 4\,\left( -3 + 2\,c \right) \,x - 6\,\left( -2 + 3\,c \right) \,x^2 +
        20\,c\,x^3 + 5\,c^2\,x^4}{3\,{\left( -1 + 2\,x + c\,x^2 \right)
        }^2}}} \ee \esub
\subsection*{Case IV: $\bs{\alpha = 2,\, \beta =-2, \, \gamma = 2, \, \delta = 0}$}
We get three solutions of \eqref{Painleve 6}. The first two \bsub\be
P(x)& =& \frac{x\,\left( -6 + c\,\left( -2 + x \right)  - 5\,x +
      4\,\left( -2 + x \right) \,\log (2 - 2\,x) \right) }{-9 - 6\,x + 2\,x^2 +
    c\,\left( -3 + 2\,x \right)  + 4\,\left( -3 + 2\,x \right) \,\log (2 - 2\,x)}\\
\nonumber & &\\ P(x)& =&\frac{-4 + \left( -30 + 4\,c \right) \,x +
\left( 13 + 2\,c \right) \,x^2 -
    8\,x\,\left( 2 + x \right) \,\log (x)}{-27 + 2\,x + 4\,x^2 + c\,\left( 2 + 4\,x \right)  -
    8\,\left( 1 + 2\,x \right) \,\log (x)} \ee\esub are
    unacceptable, since
they give again $\rho \rightarrow \frac{5}{3}$, and the third: \be
P(x)=\frac{1 - 2\,\left( 3 + 4\,c \right) \,x + 12\,c\,x^2 +
    2\,x\,\left( -2 + 3\,x \right) \,\log \left( \frac{x}{1 - x} \right) }{-4 +
    c\,\left( -4 + 8\,x \right)  + \left( -2 + 4\,x \right) \,
     \log \left( \frac{x}{1 - x} \right) } \ee which makes \eqref{contact}
\be\label{u 2.4}  u(\rho )= \frac{A(x)}{B(x)}, &&
 \rho =\frac{C(x)}{D(x)}\ee with \bsub\be A(x)& = & 4\,x\,\left( 1 - 3\,x + 2\,x^2 \right)\,
  \log^2 \left( \frac{x}{1 - x} \right)\nonumber\\
&&  +4\,\left( -1 + x \right) \,x\,\left( -1 + c\,\left( -4 + 8\,x
\right)  \right)\,
  \log \left( \frac{x}{1 - x} \right)\nonumber \\
 && + 4\,\left( -1 + 2\,c + 4\,c^2 \right) \,x - 8\,c\,\left( 1 + 6\,c \right)
  \,x^2 + 32\,c^2\,x^3+2\\ && \nonumber\\
  B(x) &= &3\,\left( 2 + 2\,c - 4\,c\,x +
      \left( -1 + 2\,x \right) \,\log (1 - x) + \log (x) - 2\,x\,\log (x) \right)^2\nonumber \\
 &&x\,\left( -1 + x \right) \\ && \nonumber \\
 C(x) & = & x\,\left( -5 + 17\,x - 24\,x^2 + 12\,x^3 \right)
 \log^2{\frac{x}{1-x}} \nonumber \\
 &&+4\,\left( -1 + x \right) \,x\,\left( 3 - 6\,x + c\,
 \left( 5 - 12\,x + 12\,x^2 \right)  \right)\log{\frac{x}{1-x}} \nonumber \\
 &&+2\,\left( -1 - 2\,\left( 3 + 6\,c + 5\,c^2 \right) \,x +
    \left( 6 + 36\,c + 34\,c^2 \right) \,x^2 - 24\,c\,\left( 1 + 2\,c \right) \,x^3 +
    24\,c^2\,x^4 \right)\nonumber \\ &&  \\
    D(x) & = &3\,\left( 2 + 2\,c - 4\,c\,x +
      \left( -1 + 2\,x \right) \,\log (1 - x) + \log (x) - 2\,x\,\log (x) \right)^2\nonumber \\
 &&x\,\left( -1 + x \right)
\ee\esub

\subsection*{Case V: $\bs{\alpha = 8,\, \beta =0, \, \gamma = 0, \, \delta = 0}$}
It cannot be linearized.
\subsection*{Case VI: $\bs{\alpha = \frac{1}{2}, \, \beta =-\frac{9}{2}, \, \gamma =
\frac{1}{2}, \, \delta = \frac{1}{2}}$} We get one solution \be
P(x)=\frac{-2 + 3\,x + 2\,c\,x^3}{-1 + 2\,x + 2\,c\,x^2}\ee which
gives \eqref{u 2.3} again.
\subsection*{Case VII: $\bs{\alpha = \frac{1}{2}, \, \beta =-\frac{1}{2}, \, \gamma =
\frac{9}{2}, \, \delta = \frac{1}{2}}$} We get one solution \be
P(x)=\frac{x\,\left( -1 - 4\,c\,x + 2\,c\,x^2 \right) }{-1 + 2\,x +
2\,c\,x^2}\ee  which  gives \eqref{u 2.3} again.
\subsection*{Case VIII: $\bs{\alpha = \frac{9}{2}, \, \beta
=-\frac{1}{2}, \, \gamma = \frac{1}{2}, \, \delta =- \frac{3}{2}}$}
We get three solutions of \eqref{Painleve 6}.

One of which is  identical to Case VI, and two more \bsub\be P(x) &
= & \frac{x\,\left( 5 + 2\,c\,{\left( -1 + x \right) }^2 - 4\,x -
4\,x^2 +
      6\,{\left( -1 + x \right) }^2\,\log (-1 + x) \right) }{3\,
    \left( 5 + 2\,c\,{\left( -1 + x \right) }^2 - 4\,x - 4\,x^2 + 2\,x^3 +
      6\,{\left( -1 + x \right) }^2\,\log (-1 + x) \right) }\nonumber \\ &&
       \\
P(x) & = & \frac{x\,\left( -3 + 2\,\left( -9 + 2\,c \right) \,x +
2\,\left( 6 + c \right) \,x^2 -
      6\,x\,\left( 2 + x \right) \,\log (x) \right) }{3\,
    \left( 1 - 6\,x + 2\,c\,x^2 + 2\,x^3 - 6\,x^2\,\log (x) \right) }\ee\esub
which are again unacceptable, since they give $\rho \rightarrow
\frac{5}{3}$.
\subsection*{Case IX: $\bs{\alpha =
2,\, \beta =0, \, \gamma = 0, \, \delta = 0} $} We get one solution
\be P(x)=\frac{\left( 1 + 2\,c \right) \,x^2}{-1 + 2\,x +
2\,c\,x^2}\ee which again leads to \eqref{u 2.3}.
\subsection*{Case X: $\bs{\alpha =
\frac{1}{2}, \, \beta =-\frac{1}{2}, \, \gamma = \frac{1}{2}, \,
\delta = -\frac{3}{2}}$} We get four solutions of \eqref{Painleve
6}. The first \be P(x)=\frac{x + 4\,c\,x^2 - 2\,c\,x^3}{-1 + 2\,x +
2\,c\,x^2} \ee gives \eqref{u 2.3}, the second \be P(x)=\frac{ 3 -
6\,\left( 1 + 2\,c \right) \,x + 12\,c\,x^2 +
      6\,\left( -1 + x \right) \,x\,\log \left( \frac{x}{1 - x} \right)   }{-4 +
    c\,\left( -4 + 8\,x \right)  + \left( -2 + 4\,x \right) \,
\log \left( \frac{x}{1 - x} \right) }\ee gives \eqref{u 2.4} and two
more \be P(x) & =&  \frac{1 - \left( 2 + c \right) \,x + x\,\log
(x)}{c + x - \log (x)} \\ &&\nonumber \\ P(x)&=&- \frac{c\,\left( -2
+ x \right)  - 2\,x + \left( -2 + x \right) \,\log (-1 + x)}
    {c + x + \log (-1 + x)} \ee which are degenerate, since they give $\rho
\rightarrow \frac{5}{3}$.

The conclusion of the above analysis is that we have two particular
solutions for the function $u(\rho)$, i.e. \eqref{u 2.3} and
\eqref{u 2.4}.

In order to write down the line element that corresponds to \eqref{u
2.3} we change the parameter $x$ and the constant $c$ to the values
\be\label{x to xi} \begin{aligned} x\rightarrow\frac{1-cosh(\xi)}{2}
&& c \rightarrow \frac{-1}{\lambda+1} \end{aligned} \ee

Gathering all the pieces we arrive at the final form of the metric:
\be\label{final G4 metric} ds^2=\kappa^2\,\biggl(
A(\xi)\,(-d\xi^2+dx^2)+B(\xi)\,e^{-2\,x}\,dy^2+2\,C(\xi)\,e^{-x}\,dy\,dz+C(\xi)\,dz^2
\biggr) \ee where \be\label{coef G4 metric}
\begin{aligned}
A(\xi)& = \frac{1}{4}\,(\cosh2\,\xi+4\,\lambda\,\cosh\xi+3) \\
B(\xi)& =
\frac{\cosh4\,\xi+8\,\lambda\,\cosh3\,\xi+28\,\cosh2\,\xi+56\,\lambda\,\cosh\xi+
32\,\lambda^2+3} {2\,(\cosh2\,\xi+4\,\lambda\,\cosh\xi+3)} \\
C(\xi) & =
\frac{16\,(1-\lambda^2)\,\sinh^2\xi}{\cosh2\,\xi+4\,\lambda\,\cosh\xi+3}
\end{aligned} \ee
with $-1<\lambda<1$ for the metric to have signature $(-+++)$.

The metric (\ref{final G4 metric}) admits a fourth killing vector
\be\label{kil of final G4}
\xi_4=-y\,\partial_x+\left(\frac{1}{8}\,e^{2\,x}-\frac{1}{2}\,y^2\right)\,\partial_y-
\frac{1}{4}\,e^x\partial_z \ee

which produces with (\ref{killing}) the following table of
(non-vanishing) commutators: \begin{equation}\label{commu2}
[\xi_1,\xi_3]=\xi_1,\,[\xi_1,\xi_4]=-\xi_3,\,[\xi_3,\xi_4]=\xi_4
\end{equation}

The existence of a fourth Killing field implies that this geometry
is an LRS space time ( see \cite{EM} where all LRS Bianchi
geometries are characterized )

The isotropy group inferred from the above algebra (see the last
commutator) is a $G_1$ spatial rotation.

The line element (\ref{final G4 metric}) is a two-parametric family
and is one of the Kinnersley vacuum solutions \cite{Kinner}.  One
way we can be assured that the constants $\kappa,\lambda$ are indeed
essential, is to consider the following three scalars, constructed
out of the Riemmann tensor and its first and second covariant
derivatives:

\be\label{scalars} Q_1=R^{ABCD}\,R_{ABCD},\qquad
Q_2=Q_{1\,;\,A}^{\ph{abcd};\,A}, \qquad
Q_3=Q_{1\,;A}\,Q_1^{\ph{a};\,A} \ee

( here capital Latin letters stand for world space time indices,
ranging over the values 0,1,2,3 )

The determinant of the Wronskian matrix $\frac{\partial
(Q_1,Q_2,Q_3)}{\partial(\xi,\kappa,\lambda)}$ can been seen to be
non zero. Therefore, $(\kappa,\lambda)$ (and of course $\xi$) can be
in principle expressed as functions of (\ref{scalars}), and are thus
essential.

The line element that corresponds to solution \eqref{u 2.4}, is
quite interesting because as it can be seen leads to an imaginary
lapse function, implying a Euclidean signature for the metric
tensor. To get a manageable form of this metric we change the
parameter $x$ and the constant $c$ to the values \be\label{x to xi
2}
\begin{aligned} x\rightarrow \arctan e^{- {\mu }^2\,\cos^2 \xi   }
&& c \rightarrow -\mu^2
\end{aligned} \ee
thus arriving at the line element \be\label{final Euclid metric}
ds^2=\kappa^2\,\biggl(
A(\xi)\,d\xi^2+B(\xi)\,dx^2+\frac{e^{-2\,x}}{B(\xi)}\,dy^2+2\,C(\xi)\,e^{-x}\,dy\,dz
+D(\xi)\,dz^2 \biggr) \ee where \be\label{coef Euclid metric}
\begin{aligned}
A(\xi)& = \frac{1}{2}\,{\mu }^4\,\sech^2({\mu }^2\,\cos^2 \xi)
\,\sin^2 2\,\xi\,
    \left( 1 + {\mu }^2\,\sin^2 \xi \,\tanh ({\mu }^2\,\cos^2 \xi ) \right)  \\
B(\xi)& = \frac{1}{2}\,
    \left( 1 + {\mu }^2\,\sin^2 \xi \,\tanh ({\mu }^2\,\cos^2 \xi ) \right) \\
C(\xi) & = \frac{\sech^2(\mu^2\,\cos^2\xi)\,\left( 2\,
\mu^2\,\sin^2\xi-\sinh(2\,\mu^2\cos^2 \xi)\right)}{1 + {\mu
}^2\,\sin^2 \xi \,\tanh ({\mu }^2\,\cos^2 \xi )} \\
D(\xi) & =\frac{\sech^2(\mu^2\,\cos^2\xi)\,\left(
1+\cosh(2\,\mu^2\cos^2 \xi)+2\, \mu^4\,\sin^4\xi \right)}{1 + {\mu
}^2\,\sin^2 \xi \,\tanh ({\mu }^2\,\cos^2 \xi )}
\end{aligned} \ee
with $(\mu,\xi) \in \mathds{R}$.

The essential nature of the constants $\kappa,\mu$, is secured by
the fact that the determinant of the Wronskian matrix
$\frac{\partial (Q_1,Q_2,Q_3)}{\partial(\xi,\kappa,\mu)}$ for the
corresponding curvature scalars \eqref{scalars} is again nonzero.
Two other interesting feature of this metric are: (a) the fact that
it \emph{does not} admit any other killing fields besides the
initially assumed \eqref{killing}, i.e. is a pure $G_3$ geometry and
(b) that it is \emph{not} either \emph{self-dual} or \emph{anti
self-dual}.

At this stage we have extracted from solution \eqref{y equation} as
much information as we possibly could, regarding its particular
solutions. The conclusion is that the only line element, in terms of
elementary functions, with Lorentzian signature is \eqref{final G4
metric} which admits a $G_4$ symmetry. This solution, as already
mentioned, was first found in \cite{Kinner} during the search of all
Petrov Type D metrics which by no doubt is a completely different
approach from ours. On the other hand the way we reproduced this
solution is by searching for particular solutions of the
Painlev\'{e} equation, a purely mathematical way of viewing, in
which no prior assumption of symmetry has been adopted. We find it
interesting to retrieve the same line element by previously assuming
the existence of a fourth killing field. This is what we will do in
the next section, exploiting  a "hidden" symmetry of the Bianchi
Type III 3-Geometry.

\section{Derivation of the Lorentzian solution assuming a $G_4$ symmetry}

In \cite{ChrTer} we were not able to obtain the full solution
\eqref{final u III}, thus we presented  only an one-parameter family
of Bianchi Type III Cosmology. The line element is of the general
form (\ref{lin el Na=0}) with an overall essential constant
$\kappa^2$: \be\label{metric u4 not con} ds^2 & = &\kappa^2\,\left(
-\frac{e^{2\,\xi}\,(e^{2\,\xi}+1)}{4\,(2\,e^{2\,\xi}+1)}\,d\xi^2+
\frac{e\,^\xi}{4}\,cosh\xi\,dx^2+e^{-2x+\xi}(cosh2\xi+2)\,sech\xi\,dy^2
\right.
\nonumber \\
\nonumber \\
& &
\left.\ph{-\frac{e^{2\,\xi}\,(e^{2\,\xi}+1)}{4\,(2\,e^{2\,\xi}+1)}\,}
+\,e^\xi\,sech\xi\,dz^2+2e^{-x+\xi}\,sech\xi\,dy\,dz \right) \ee

This metric admits, besides the three killing fields
(\ref{killing}), yet another one, namely: \be\label{newkil}
\xi_4=-y\,\partial_x+\frac{e^{2\,x}-8\,y^2}{16}\,\partial_y-\frac{1}{8}\,e^x\,\partial_z
\ee The commutator table of their algebra is: \be\label{comm table}
\begin{aligned}
\bigl[\xi_1,\xi_2\bigr]&=0 & \bigl[\xi_1,\xi_3\bigr]&=\xi_1 &
\bigl[\xi_1,\xi_4\bigr]&=-\xi_3 \\
\bigl[\xi_2,\xi_3\bigr]&=0 & \bigl[\xi_2,\xi_4\bigr]&=0 &
\bigl[\xi_3,\xi_4\bigr]&=\xi_4
\end{aligned}
\ee

An  equivalent form of the Type III member of the known
Ellis-MacCallum family of solutions \cite{MacBook},\cite{Wainr},
also retrieved in \cite{ChrTer}, reads: \be\label{macdiag}
ds^2=\lambda^2\,\left(-\frac{e^{3\,t}}{e^{t}-1}\,dt^2
+e^{{2\,t}}\,dx^2+e^{2\,t-2x}\,dy^2+(1-e^{-\,t})\,dz^2\right)\ee
which again admits a fourth killing field: \be\label{newkil Mac}
\eta=-y\,\partial_x+\frac{e^{2\,x}-y^2}{2}\,\partial_y \ee The
interesting thing is that both quadruplets
$\left(\xi_1,\xi_2,\xi_3,\xi_4\right)$,
$\left(\xi_1,\xi_2,\xi_3,\eta\right)$ span the same algebra. Yet the
two line elements (\ref{metric u4 not con}) and (\ref{macdiag}) are
inequivalent, i.e. we cannot arrive from one to the other by a
coordinate transformation. This can be easily seen since for metric
(\ref{macdiag}) we have an invariant relation of the form:

\be\label{invar1}\frac{18\,Q_1^7}{(Q_2\,Q_1-Q_1^{\ph{1};\,A}\,Q_{1;\,A})^3}=
\lambda^2,\,Q_1=R^{KLMN}R_{KLMN},\,Q_2=\Box{R^{KLMN}R_{KLMN}} \ee
where capital Latin letters denote space-time indices ranging in the
interval (0-3), the semicolon stands for covariant differentiation,
and the
 $\Box$ for the covariant D'Alebertian. This relation, being a constant scalar constructed out of the
intrinsic geometry (the Riemmann tensor and its covariant
derivatives), characterizes, along with many others that can be
found, this metric: It will be valid for any equivalent, under
general coordinate transformations, form of (\ref{macdiag}). But for
 metric (\ref{metric u4 not con}) the left hand side of
(\ref{invar1}) does not equal $\kappa^2$, so the two metrics are
inequivalent.

The way that the two solutions were found, belonging in different
"branches" of the solution space, and without any assumption of
extra symmetry, suggests that the existence of the fourth killing
field may not be a mere coincidence: Indeed the very existence of a
3-space with a Type III symmetry group implies the existence of a
$G_4$ action, a thing that is not so common.

In order to  see this  we consider an arbitrary hypersurface with
$t=t_o$ of the space-time (\ref{lin el Na=0}): \be\label{spatial
metric} dl^2=
\gamma_{\alpha\beta}\,\sigma^\alpha_{\ph{a}i}\,\sigma^\beta_{\ph{a}j}\,dx^i\,dx^j
\ee where the scale factor matrix is given by \be
\gamma_{\alpha\beta}=\left(
\begin{matrix}
e^{{u_1} + 2\,{u_4}} & e^{{u_1} + {u_2} + {u_4}} & 0 \cr e^{{u_1} +
{u_2} + {u_4}} & e^{{u_1} + 2\,{u_2}}\,t_o & 0 \cr 0 & 0 & e^{{u_1}}
\end{matrix}
\right) \ee and the functions $u_i$ are evaluated at $t=t_o$.
Solving the killing equation for this line element, we surprisingly
find that, besides the three killing fields (\ref{killing}), there
exists a fourth:

\be\label{4 killing} \zeta=2\,y\,\partial_x+
\left(y^2-\frac{e^{-2\,u_4}\,t_o}{4\,(t_o-1)}\,e^{2\,x}\right)\,\partial_y+
\frac{e^{-u_2-u_4}}{2\,(t_o-1)}\,e^x\,\partial_z \ee This field
would exist even if we have filled out the zero entries of the scale
factor matrix $\gamma_{\alpha\beta}$ using the constant
Automorphisms. The qualitative difference between (\ref{4 killing})
and (\ref{killing}) is that the  components of the former depend on
$\gamma_{\alpha\beta}$, and thus this vector is not form invariant
like the latter.

If we wish this 3-dimensional killing field, to be promoted to an
isometry of space-time, we first need to prolong it by adding a time
component say of the form $f(t,x,y,z)\,\frac{\partial}{\partial t}$.
Omitting  the calculational details, the result is that $f(t,x,y,z)$
should be zero (trivial prolongation) and the components of $\zeta$
must be independent of the slice parametrization. Thus the
derivative with respect to $t_o$, for every component of $\zeta$, is
zero, i.e. \be \left\{
\begin{split}\frac{d}{dt_o}\left(\frac{e^{-2\,u_4(t_o)}}{4\,(t_o-1)}\,t_o\right)&
=& 0
\\
\frac{d}{dt_o}\left(\frac{e^{-u_2(t_o)-\,u_4(t_o)}}{2\,(t_o-1)}\right)&
=& 0 \end{split}\right.\Rightarrow \left\{\begin{matrix} u_2(t)& = &
k_2-\frac{1}{2}\,\ln(t^2-t) \\
\\ u_4(t)&=&k_4-\frac{1}{2}\,\ln\left(\frac{t-1}{t}\right)
\end{matrix}\right.\ee
Inserting these values into the scale factor matrix
$\gamma_{\alpha\beta}$ we have: \be\label{gamma G4}
\gamma_{\alpha\beta} = \left(\begin{matrix} \frac{t}{t-1}\,e^{u_1} &
\frac{1}{t-1}\,e^{u_1} & 0
\\
\frac{1}{t-1}\,e^{u_1} & \frac{1}{t-1}\,e^{u_1} & 0
\\
0 & 0 & e^{u_1} \end{matrix} \right) \ee where we have absorbed the
constants $(k_2,k_4)$ with a re-scaling of the coordinates
$(y,z)\rightarrow(e^{-k_4}\,y,e^{-k_2}\,z)$. It is easy to see that
this scale factor matrix is positive definite in the interval
$t\in(1,+\infty)$. Using (\ref{gamma G4}) in the line element
(\ref{lin el Na=0}) we are now ready to solve the vacuum Einstein
Field Equations (EFE) $R_{AB}=0,\, \left\{A,B\right\}\,\in
\,\left\{0,1,2,3\right\}$.

From the $R_{00}=0$ component we deduce: \be\label{lapse G4}
N^2=\frac{\left(3\,(t-1)\,u_1'-2\right)\,u_1'}{4\,(4\,t-3)}\,e^{u_1}
\ee Inserting this value of the lapse function, into the rest of
EFE, we arrive at a \emph{single} second order ODE for the function
$u_1(t)$: \be\label{final u1 G4}
u_1''=-\frac{6\,t\,(t-1)^2\,{u_1'}^2+(10\,t^2+13\,t-3)\,u_1'+8\,t-5}{4\,t^2-7\,t+3}
\,u_1' \ee which is an Abel equation for the function
$w(t):=u_1'(t)$. The general solution to this equation (see
Appendix) for the function $w(t)$ is given in implicit form:
\be\label{implicit sol}
G(t,w):=\frac{\left(3\,(t-1)\,w-2\right)^3\,(t-1)^3\,w}{\left(2\,(t-1)^2\,(t-3)\,w^2
-2\,(t^2-4\,t+3)\,w-1\right)^2\,(4\,t-3)} =
\text{\emph{constant}}\ee One possible parametrization of
$G(t,w)=const$ is \be\label{param} w & = &
\frac{64\,(1-\lambda^2)\,(\lambda+\cosh\xi)\,\sinh^4\xi}
{(\cosh3\,\xi-9\,\cosh\xi-8\,\lambda)\,(\cosh
2\,\xi+4\,\lambda\,\cosh\xi+3)^2}
\nonumber \\ \nonumber \\
t & = &
\frac{\cosh4\,\xi+8\,\lambda\,(\cosh3\,\xi+7\,\cosh\xi)+28\,\cosh2\,\xi+32\,\lambda^2+3}
{32\,(1-\lambda^2)\,\sinh^2\xi}\ee yielding  \be\label{G constant}
G(t,w)=\frac{1-\lambda^2}{12\,\lambda^2} \ee Of course, as it is
evident from (\ref{param}), this parametrization is not valid for
$\lambda=\pm1$, but then, from (\ref{G constant}) we deduce that
$G=0\Rightarrow u_1'=\frac{2}{3\,(t-1)}$ which is unacceptable since
it makes the lapse function (\ref{lapse G4}) zero. Another breakdown
of (\ref{param}), is when the denominator of $G(t,w)$ vanishes.
However, this is the case that corresponds to the line element
(\ref{metric u4 not con}), i.e. emerges from a special solution of
(\ref{final u1 G4}). Gathering all the pieces we arrive at the line
element \eqref{final G4 metric}.

\section{Conclusions}
We have seen how the Automorphisms of Type III Geometry can be used
as symmetries of the corresponding EFE's, in order to reduce the
degree of these equations, and ultimately integrate them in full.
The solution space of the differential equations, is seen to be
naturally partitioned in three disconnected components: One occupied
by the Type III member of the known Ellis-MacCallum family
(\ref{macdiag}), another described  by the equation (\ref{final u
III}) which is fully integrated by the parametrization
(\ref{solution y}) and a piece occupied by the known Siklos
solution, an equivalent form of which is

\be\label{metric u4=con}
ds^2=-\lambda^2\,d\xi^2+\frac{\xi^2}{4}\,dx^2+e^{-2x}\xi^{4\lambda}\,dy^2
+\frac{\lambda-1}{2\lambda-1}\,dz^2+2e^{-x}\xi^{2\lambda}\,dy\,dz
\ee

This line element can be obtained from (\ref{gamma u}), for the
special case in which
$u_3:=\frac{\gamma_{11}\,\gamma_{22}}{\gamma_{12}^2}=cont$; $u_3$ in
this paper is, by a choice of time "gauge", taken to be the term
$\frac{3\,\rho-3}{3\,\rho-5}$ in $\gamma_{\alpha\beta}$.

But at the level of the geometry, a unification of the first two
branches might be achieved, at the expense of complete mathematical
rigor. Consider the particular solution to the Painlev\'{e} equation
\eqref{Painleve 6} which corresponds to the two-parameter family
\eqref{final G4 metric} (Kinnersley solution): This solution can be
seen to incorporate both \eqref{metric u4 not con} found in
\cite{ChrTer}, for an admissible choice of $\lambda$ and the Type
III member of Ellis-MacCallum metric \eqref{macdiag} for the
marginal value $\lambda=1$ (since this is the only value for which
the invariant relation \eqref{invar1} is satisfied by the
two-parameter family). However, for $\lambda=1$ the coefficient of
$dz^2$ is vanishing along with the cross term $dy\,dz$. In order to
avoid this incompatibility (zero eigenvalues) one may first employ
the transformation \be z\rightarrow \frac{z}{2\,\sqrt{1-\lambda^2}},
& y\rightarrow \frac{y}{2}, & \xi \rightarrow
\arccosh\frac{e^{t/2}}{\sqrt{2}} \ee which cancels the $1-\lambda^2$
factor in the $dz^2$ term, while maintains a $\sqrt{1-\lambda^2}$ in
the cross term; now putting $\lambda=1$ results in the diagonal
solution \eqref{macdiag}.

The fact that this solution admits a $G_4$ symmetry group, is not
due to a prior assumption of this symmetry, but emerged out of the
particular nature of the solution of the Painlev\'{e} equation. Of
course, as shown in detail in section 3, the same solution can be
retrieved by first assuming the existence of the fourth killing
field. It is interesting that the form of this field is almost
dictated by the unknown existence of a fourth Killing field on the
slices: for any line element of the form \eqref{spatial metric},
there is a fourth Killing field \eqref{4 killing}. Promoting it to a
space-time Killing field we arrive, through the solution of
\eqref{final u1 G4}, to the aforementioned solution.

On the other hand, the solution \eqref{final Euclid metric}, which
is of Euclidean signature, admits no extra Killing field and is thus
a genuine $G_{3}$ geometry. It also contains 2 essential parameters,
which implies that the number of essential parameters is not
associated to the dimension of the isometry group.

The number of essential constants contained in the solution
described by \eqref{Painleve 6} is 3, as expected for the Type III
vacuum  Cosmology: two implicit in the Painlev\'{e} transcendent
plus one multiplicative constant in front of the line element
\eqref{lin el Na=0}, owing to the fact that this line element admits
no homothetic Killing field.

 We believe that the present work constitutes an adequate explanation for the scattered
 occurrence of various Painlev\'{e} transcendents in the literature
 on Bianchi solutions and  we expect to be able to present the corresponding
general solutions, along with the known ones, for all lower Bianchi
Types (I-VII), in a forthcoming publication.

\textbf{Acknowledgements}\\
 The authors are  indebted to prof. M.A.H. MacCallum
who, in a private communication pointed out that the solution
\eqref{final G4 metric} belongs to Kinnersley's family and is not
new. The project is co-funded by the European Social Fund and
National Resources - (EPEAEK II) PYTHAGORAS II .

\newpage
\section*{Appendix}
\appendix
\setcounter{section}{1} \setcounter{equation}{0}
 As we already mention the equation
(\ref{final u1 G4}) is an Abel equation of the first kind for the
function $w(t):=u_1'(t)$. Since there is no general method to obtain
the solution of such an equation we have tried a different approach,
based on the application of Prelle-Singer algorithm \cite{PreSin}.

In brief the Prelle-Singer algorithm (actually semi-algorithm) is a
method for finding integrating factors for first order differential
equations of the form: \be\label{1st order}
y'=\frac{dy}{dx}=\frac{M(x,y)}{N(x,y)} \ee where $M(x,y)$ and
$N(x,y)$ are polynomials with coefficients in the field of complex
numbers, $\mathbb{C}$. In \cite{PreSin}, Prelle and Singer proved
that, if an elementary first integral of (\ref{1st order}) exists,
it is possible to find an integrating factor $\mu$ for this
equation, i.e. \be \frac{\partial(\mu\,N(x,y))}{\partial\,x}+
\frac{\partial(\mu\,M(x,y))}{\partial\,y}=0 \ee For this purpose
they defined the operator \be\label{D operator}
\mathcal{D}=N\,\frac{\partial}{\partial
x}+M\,\frac{\partial}{\partial y} \ee and the Darboux polynomials
$f_i$, i.e. irreducible polynomials that obey \be\label{Darboux}
\mathcal{D}f_i=f_i\,g_i \ee for some polynomial $g_i$. Then, to find
the integrating factor $\mu$, one has to choose a degree
$N_D=\left(\text{degree}(x),\text{degree}(y)\right)$ for the
polynomials $f_i$, calculate them and if a relation of the form
\be\label{Prelle eq} \sum_{i=1}^k n_i\,g_i=-\left(\frac{\partial
N}{\partial x}+\frac{\partial M}{\partial y}\right) \ee is
satisfied, for some non-zero rational numbers $n_i$, then the
integrating factor $\mu$, is given by \be\label{mu}
\mu=\prod_{i=1}^k f_i^{n_i} \ee Unfortunately the method can not
define the value of the degree $N_D$ of the polynomials, that's why
it is a semi-algorithmic approach. Nevertheless it guarantees, that
if an elementary first integral exists, it can be found.

Returning now to equation (\ref{final u1 G4}), in order to bring it
in a form suitable for the Prelle-Singer method, we apply the
transformation $w(t)=\frac{2}{3\,(t-1)\,y(t)}$ resulting to
\be\label{Pre w}
y'=\frac{2\,(y-1)\,(6\,t\,y-4\,t-3\,y)}{3\,y\,(4\,t^2-7\,t+3)} \ee

The Darboux polynomials $f_i$ of degree $N_D=(1,1)$ and their
corresponding polynomials $g_i$ for the operator \be
\mathcal{D}=3\,y\,(4\,t^2-7\,t+3)\,\frac{\partial}{\partial t}+
2\,(y-1)\,(6\,t\,y-4\,t-3\,y)\,\frac{\partial}{\partial y} \nonumber
\ee are \be\label{Darboux G4} \begin{aligned} f_1 & =4\,t-3 & g_1 &
=12\,t\,y-12 \\
f_2 & = t-1 & g_2 &=12\,t\,y-9\,y \\
f_3 & = 4\,t-3\,y & g_3 & = 12\,t\,y-6\,y-6 \\
f_4 & =-t\,y+y+t-1 & g_4 & =24\,t\,y-15\,y-8\,t
\end{aligned} \ee
Inserting this polynomials in (\ref{Prelle eq}), we can compute the
numbers $n_i$, which read \be
\left(n_1,n_2,n_3,n_4\right)=\left(-\frac{1}{2},\frac{1}{2},1,-\frac{5}{2}\right)
\ee yielding an integrating factor of the form \be\label{integr mu}
\mu=\frac{(4\,t-3\,y)\,\sqrt{t-1}}{(t+y-t\,y-1)^{5/2}\,\sqrt{4\,t-3}}
\ee At this stage it is a trivial task to integrate (\ref{final u1
G4}) for $w(t)$ to obtain \be
G(t,w):=\frac{\left(3\,(t-1)\,w-2\right)^3\,(t-1)^3\,w}{\left(2\,(t-1)^2\,(t-3)\,w^2
-2\,(t^2-4\,t+3)\,w-1\right)^2\,(4\,t-3)} =
\text{\emph{constant}}\nonumber\ee


\end{document}